# Terminologies, modèles de données archéologiques et thésaurus documentaires : réflexions à partir d'une typologie de céramique

*« Tout le problème vient en fin de compte du fait qu'on tente d'appliquer un procédé graphique simple, le tableau, à l'étude des "symboles" dans les cultures orales. Il est douteux que soumettre les mots et leurs significations à un tel réductionnisme puisse être de quelque profit, même si certains ensembles peuvent mieux que d'autres s'accommoder de ce genre de traitement. Car, par ces simplifications, on produit un ordre superficiel qui est évidemment bien plus le reflet de la structure matricielle utilisée que de la structure de l'esprit (ou d'un esprit) humain, ce qui donne ces analogies générales et grossières communes à toutes les constructions de ce style. »*

- Jack Goody, *La raison graphique : la domestication de la pensée sauvage*, 1979, p. 132.

**Sébastien DUROST\*, Guillaume REICH\*\*, Jean-Pierre GIRARD\*\*\***

\* Responsable de la cellule éditoriale et de la stratégie numérique, Bibracte - BIBRACTE EPCC - Centre archéologique européen, F-58370, Glux-en-Glenne / s.durost@bibracte.fr
\*\* Ingénieur de recherche, Maison des Sciences de l'Homme et de l'Environnement Claude-Nicolas Ledoux - UAR 3124, CNRS, Université Bourgogne Franche-Comté, F-25000, Besançon / dr.guillaume.reich@gmail.com
\*\*\* Chercheur associé (données ouvertes) - Archéorient - UMR 5133 - Environnements et sociétés de l'Orient ancien - Maison de l'Orient et de la Méditerranée - Jean Pouilloux – F-69365, Lyon / jean-pierre.girard@mom.fr

**Résumé**
Les projets *HyperThésau* et *Bibracte numérique* ont donné naissance à un travail collectif centré sur l'usage du vocabulaire comme axe de l'interopérabilité des données archéologiques tout au long de leur cycle de vie. Pour ce faire, l'usage de la forme normalisée du thésaurus – via la plateforme *Opentheso* – fournit un outil déjà adapté au « web des données ». Néanmoins, son usage a rapidement soulevé la question des différents paradigmes présidant à l'élaboration d'un vocabulaire particulier à chaque (groupe de) scientifique(s). La norme ISO 25964, conçue pour la gestion et l'interopérabilité des langages documentaires, se révèle assez souple pour permettre de comparer et relier différents « points de vue » de recherche scientifique ou documentaire, mais la mise en cohérence de ces points de vue au moyen d'alignements permettant leur interopérabilité nécessite d'interfacer différentes granularités sémantiques : le signalement des recherches, les descriptions des données « primaires », une passerelle ou « pivot » entre les deux, en mettant en œuvre une méthodologie de coopération régulée. Les défis qui restent à relever sur cette voie n'empêchent pas l'outil thésaurus d'être, d'ores et déjà, un support adapté à une interopérabilité « humain-machine-humain » complète, mise au point dans le cadre du projet *Bibracte Ville Ouverte* à partir d'un corpus issu d'une recherche sur la céramique de ce site archéologique.

**Mots-clés**
thésaurus ; usages ; interopérabilité ; sémantique/terminologies ; preuve de concept ; archéologie

**Summary**
The *HyperThésau* and *Bibracte numérique* projects have given rise to a collective effort centred on the use of vocabulary as a means of ensuring the interoperability of archaeological data throughout its life cycle. To this end, the use of the standardised form of the thesaurus – via the *Opentheso* platform – provides a tool that is already adapted to the Linked Data. Nevertheless, its use quickly raised the question of the different paradigms presiding over the elaboration of a specific vocabulary by each (group of) scientist(s). The ISO 25964 standard – designed for the management and interoperability of indexing languages – is flexible enough to permit the comparison and linking of different scientific or documentary "points of view". Their coherence through interoperability alignments nevertheless requires to interface different semantic granularities: search reporting, the description of raw data, a gateway or "pivot" between the two, by using a regulated cooperation methodology. The challenges that remain to be met on this path do not prevent the thesaurus tool from already being a suitable support for a complete "human-to-machine-to-human" interoperability, developed within the framework of the *Bibracte Ville Ouverte* project and exemplified through a research on the ceramics of that archaeological site.

Cet article est le premier d'une série – émanant des mêmes auteurs principaux – qui vise à mettre en discussion la démarche pour trouver, accéder, interopérer et réutiliser des données archéologiques[1] que mènent ensemble, depuis trois ans, quelques acteurs et institutions du domaine de l'archéologie, à travers une série de projets dont les avancées sont rythmées par l'obtention, ou non, de crédits issus des désormais incontournables appels à projets.

Les principes en ont été exprimés en 2018 dans l'argumentaire du projet *HyperThésau*[2] : « *Dans une démarche orientée "usages", une chaîne d'outils méthodologique, sémantique et technologique sera appliquée aux données de sites* [archéologiques] *urbains anciens plus ou moins massivement numérisés. En centrant la réflexion sur le rapport de l'humain-utilisateur à des "objets de savoir" archéologiques en contexte numérique, on abordera* [la question du] *"cycle de vie"* [...] *des données acquises, organisées, étudiées, archivées et partagées par des professionnels, le plus souvent* "avec les moyens du bord". *"HyperThésau" concevra, prototypera et testera en conditions réelles un mini-système d'information : un flux collaboratif scientifique associant méthodologie interdisciplinaire et expertise-métier pour l'enrichissement sémantique et la documentarisation de données archéologiques* […] *; l'articulation d'un design de base de données, d'outils logiciels et d'une architecture-système pour l'enregistrement, l'interrogation, l'enrichissement et la gestion évolutive de ces données, leur archivage pérenne dans les outils de la TGIR* Huma-Num *et le respect de leur interopérabilité* […]. » ; ils ont été mis en pratique dans le cadre, entre autres, du projet *Bibracte Numérique*[3].

Le vocabulaire y est immédiatement apparu comme le pivot de l'intercompréhension des travaux et de leurs comptes-rendus. En effet, les seules données archéologiques réellement « brutes » sont l'unité stratigraphique ou le vestige exhumé, tel l'objet physique que l'on peut directement observer. Dans la pratique ne sont ensuite manipulées que des « traces primaires », descriptions graphiques (minute dessinée ou orthophotographie, par exemple) ou textuelles. Or, il n'existe pas de norme ISO (https://www.iso.org/) pour la représentation de la donnée archéologique, ni de consensus en tenant lieu pour les logiciels d'acquisition et de gestion des données ainsi que pour le vocabulaire utilisé, alors que la perception (donc la description) d'un fait archéologique diffère selon l'usage, l'usager et l'état du savoir donné (Bruneau 1992 ; De Luca *et al.* 2016, pp. 8-10) ; si la normalisation du dessin technique en archéologie (par exemple Feugère *et al.* 1982), et en particulier en céramique (Arcelin, Rigoir 1979), est relativement consensuel, rien d'équivalent n'existe, au moins à l'échelle nationale, pour les termes de vocabulaire. « Lire » des libellés revient donc à les interpréter, selon différents paradigmes : paradigme référentiel (associer un libellé à un objet ou un concept correspondant), paradigme psychologique (associer un libellé à une représentation/image mentale) ou paradigme différentiel (mettre en relation différents libellés définis par un jeu d'identités et de différences), dans un univers sémantique dès lors à définir (Bachimont 2000, p.5 se référant à la définition des trois paradigmes de Rastier *et al.* 1994).

Pour « faire avec » cette absence de consensus normatif sur les libellés archéologiques, on peut utiliser les thésaurus documentaires dont la construction est, elle, régie par une norme (ISO-25964-1) ; celle-ci fixe des règles formelles : caractéristiques d'un concept ; structure et gestion du thésaurus ; données minimales à rassembler pour répondre à la norme. Celle-ci, en tant que telle, définit des règles et une syntaxe mais, bien qu'appliquée au langage, est totalement dépourvue de sémantique (Searle 1980 ; 1987, pp. 87-90).

---

[1] « *Findable, Accessible, Interoperable, Reusable* », d'où l'acronyme *FAIR* recouvrant la notion des *FAIR data* ou données *FAIR*. Ces principes recouvrent, dans le contexte de l'accessibilité de l'Internet, des données de la recherche et des sciences ouvertes (et de manière générale du partage et de l'ouverture des données), les manières de construire, stocker, présenter ou publier des données de manière à ce qu'elles soient *faciles à trouver*, *accessibles*, *interopérables* et *réutilisables*.

[2] Le projet *HyperThésau* (2018-2020), financé par le *LabEx Intelligences des Mondes Urbains*, prévoyait la constitution de « micro-thésaurus pivots » pour l'archéologie, issus des pratiques-métier infra-disciplinaires des équipes de recherche et alignés sur les grands référentiels du web sémantique (*Library of Congress Subject Headings*, data.bnf.fr, IdREf, etc.). L'approche envisagée reposait en outre sur la création d'une architecture informatique nouvelle (le « lac de données ») susceptible d'ingérer, stocker et analyser une grande masse de documentation, quels que soient les formats utilisés. À titre de preuve de concept, un premier micro-thésaurus *« Prospection et techniques de la géophysique »* a été publié sur la plateforme de référence *Opentheso* (Maison de l'Orient et de la Méditerranée Jean Pouilloux) : https://ark.mom.fr/ark:/76609/arcbvn2dd9cdw

[3] Depuis 2017, Bibracte EPCC (http://www.bibracte.fr/) s'est attaché à mettre en place un programme pluriannuel (2018-2021) de développement d'outils numériques pour ses différents métiers et les différentes catégories de ses usagers (archéologues, étudiants, chercheurs et grand public), dans une perspective d'approche intégrée, d'expérimentation et de partage d'expérience avec d'autres acteurs de l'archéologie. Le programme a été construit autour de quatre projets pour intégrer les possibilités offertes par les humanités numériques : diffusion de la connaissance par une médiation numérique, équipement numérique des chantiers de fouille, chaîne de production de la connaissance archéologique et prototypage d'une démarche pour trouver, accéder, interopérer et réutiliser des données archéologiques.

Les thésaurus les plus complets disponibles aujourd'hui sont produits par des documentalistes. Ils auraient vocation à devenir des points de repère pour les chercheurs, mais cela pose la question de l'intentionnalité collective de la démarche (De Munck 1998, à propos de la pensée de Searle 1980 ; 1987 ; 1990 ; 1995). Un dispositif socio-technique, quel qu'il soit, rend compte dans sa mise en œuvre de choix contraints par une adaptation-négociation entre, d'une part, les possibilités et limites de la technologie mobilisée, d'autre part, les objectifs de service portés par ses opérateurs, enfin les besoins ou demandes d'usage des utilisateurs finaux : l'organologie du dispositif influe sur les capacités des acteurs (Stiegler 1994 ; 2014). Pour la création d'*index bibliographiques* structurés par la logique formelle de la norme ISO-25964-1, toute l'attention des opérateurs a été focalisée, dans leur mise en œuvre, sur la différenciation intralinguistique (Bachimont 2000, p. 9) des libellés, ainsi transmutés en concepts non-ambigus dans un thésaurus arborescent (Maroye 2015, pp. 76-78). Ce modèle organologique des thésaurus bibliographiques est-il mobilisable pour un vocabulaire « orienté recherche » et, si oui, comment ? Cette question est l'un des enjeux du projet *Bibracte Ville Ouverte*[4].

**Des besoins sémantiques spécifiques pour l'archéologue**

L'interopérabilité des vocabulaires archéologiques créés par et pour la recherche est complexe car ces derniers sont multiples et reflètent la diversité des problématiques de recherche, leur évolution et la variabilité des corpus sollicités. L'archéologue organise et rend compte des données matérielles brutes qu'il appréhende en mobilisant des connaissances accumulées au sein de la communauté scientifique. Selon la nature de sa problématique, il sélectionne et observe plus particulièrement tel(s) ou tel(s) paramètre(s) intrinsèque(s) de son objet d'étude, qui lui semble(nt) pertinent(s) pour son travail général de modélisation de la connaissance. Il utilise les plus petites unités descriptives congruentes pour rendre compte de ce qu'il observe sur le terrain ou sur le mobilier, puis les hiérarchise en fonction de la résolution ou granularité de sa problématique initiale. Son échelle d'observation – dépendante du corpus traité – oscille entre objectivité et subjectivité, avec le souci permanent d'être à l'épreuve des faits archéologiques. La concaténation des différentes caractéristiques repérées (paradigme différentiel, *cf. supra*) lui permettent alors de comprendre et de construire un type, c'est-à-dire une unité taxinomique caractérisée par une définition univoque (textuelle et généralement associée à une/des représentation(s) iconographique(s) idoine(s)). Toutes ces informations sont synthétisées par l'archéologue dans des typo-chronologies (inscription des types sur une échelle temporelle) lui servant d'appui pour produire un discours sur les sociétés humaines passées.

Le soubassement scientifique de la recherche archéologique implique de ne jamais se départir du contexte observable. C'est tout le paradoxe de cette discipline, qui propose des modèles de représentation du monde s'appuyant sur des contextes d'étude toujours différents et des points de vue variés, c'est-à-dire des paradigmes variables selon le lieu et le moment de leur mobilisation. Cela renvoie à la question de la validité temporelle et spatiale de la définition d'un concept : l'augmentation du volume des découvertes archéologiques et l'évolution des raisonnements obligent la communauté scientifique à une sempiternelle réactualisation. À mesure que les corpus s'étoffent et que les connaissances progressent, les définitions évoluent. Certaines connaissances/notions archéologiques se précisent en granularité, d'autres sont abandonnées et de nouvelles sont créées, suivies simultanément par l'actualisation du vocabulaire en vigueur. À fins épistémologiques et historiographiques, il faut pouvoir rendre compte de l'état de la recherche ; c'est pour cette raison que les vocabulaires et typologies doivent impérativement être contextualisés, millésimés et rester durablement accessibles.

Par exemple, des céramologues de Bibracte (Barrier, Luginbühl 2021) parlant d'une céramique comme appartenant à la forme « assiette », au type « A15 » et à la catégorie « PGFINLF » mobilisent *de facto* des critères implicites. Pour eux, une catégorie céramique informe aussi bien sur la couleur de la pâte, la composition minéralogique de la terre, l'origine des matières premières entrant dans sa fabrication, le traitement thermochimique de l'argile ou le mode de façonnage de la poterie. Une forme et un type céramiques, quant à eux, les renseignent sur un contexte chrono-culturel, les données métriques de l'artefact, les différences morphologiques et la fonction de l'objet. Par le mot « assiette », ces céramologues désignent une « assiette/plat », c'est-à-dire dans leur cas une forme basse, ouverte, de hauteur généralement inférieure au quart du diamètre, permettant de servir et présenter des aliments. Par l'information « A15 », ils précisent qu'il s'agit d'un plat à paroi

---

concave, à lèvre arrondie épaissie en bandeau, et qu'il s'agit d'une imitation de la forme « R-Pomp 1 » (plat à engobe interne italien, décrit dans Goudineau 1970, pl. 1, n°1 ; Beltrán Lloris 1990, n°926 et le *Dicocer* : http://dicocer.cnrs.fr/)[5]. Ils indiquent que ses premières occurrences apparaissent vers 60 avant notre ère, deviennent plus fréquentes à partir de 50 avant notre ère et connaissent un *floruit* à partir de 30 avant notre ère jusqu'à la fin de l'époque augustéenne. La catégorie « PGFINLF », acronyme désignant une céramique à pâte grise fine et surface lissée fumigée, renseigne sur la surface – parois lissées, assez ou peu luisantes, de couleur gris foncé ou noire, présentant des types de décors variés (imprimés, imprimés à la molette, polis) –, sur la pâte – siliceuse, fine, dure, gris moyen –, sur le montage – tournage et tournassage –, sur un répertoire d'attribution – vaisselle de table et de stockage d'appoint -, sur l'origine locale et sur son contexte chronologique de production – dès avant 120 avant notre ère jusqu'au début du Iᵉʳ siècle de notre ère. Par ailleurs, toutes ces notions servant à caractériser la morphologie et la composition d'un artefact céramique renvoient ici implicitement, par un jeu de mise en abîme, à d'autres notions convoquées par le spécialiste – « fumigée », « grise », « fine », « stockage d'appoint », etc. – dont le périmètre n'a pas été systématiquement circonscrit par des critères mesurables (quantification et reproductibilité) et clairement explicités.

## Le thésaurus PACTOLS et le gestionnaire de thésaurus *Opentheso*

Dans sa pratique, l'archéologue manipule en permanence un faisceau d'indices convergents formant des définitions restituées dans des publications spécialisées dont la bibliothèque de Bibracte détient un fonds important. Le projet *Bibracte Ville Ouverte* ambitionne de créer un portail internet pour les ressources en ligne de Bibracte, valorisant le catalogue de sa bibliothèque et favorisant l'accès aux différentes formes de diffusion des résultats de son programme de recherche – données, rapports et publications. Dans une optique de mutualisation des moyens et des compétences, le *Catalogue Collectif Indexé* (*CCI*) du partenaire GDS *Frantiq* (Nouvel, Rousset 2015) s'est imposé. Seul index bibliographique national inter-établissements réservé à l'archéologie, il sera mobilisé pour exposer le catalogue de la bibliothèque de Bibracte.

L'indexation du *CCI* est organisée grâce au thésaurus PACTOLS[6] (Nouvel 2019 ; Nouvel *et al.* 2019), administré avec le gestionnaire de thésaurus multilingue et multi-hiérarchique *Opentheso*[7], diffusé en *open source*. Outil générique proposé dans la grille de services de la TGIR *Huma-Num*, ce gestionnaire offre une grande souplesse d'utilisation, facilite l'alignement entre thésaurus et permet l'interconnexion et l'alignement à tous les niveaux hiérarchiques de ses arborescences terminologiques. *Opentheso* comme PACTOLS se fondent sur la norme ISO 25964[8] (Hudon 2012), dont le premier volet a été mobilisé par Bibracte pour rendre compte de son vocabulaire spécialisé.

---

[5] Beaucoup d'études céramologiques récentes s'articulent autour de trois critères : technique (= catégories), morpho-typologie (= formes et types) et fonction. La fonction d'une céramique n'est pas entendue comme une caractéristique intrinsèque de sa morpho-typologie d'appartenance, mais relevant davantage d'appréciations complémentaires, comme la tracéologie, l'analyse des caramels alimentaires, etc. Ici, la désignation de l'assiette A15 comme une « assiette/plat » soulève une ambiguïté, puisque la fonctionnalité de la céramique est précisée et déduite à partir d'un « plat à engobe interne italien », le R-Pomp 1, décrit par ailleurs comme une « écuelle » ; le plat pouvant être vu par certains céramologues comme servant à la cuisson alors que ce n'est pas le cas de l'assiette ou de l'écuelle.

[6] Le thésaurus PACTOLS (https://pactols.frantiq.fr) est organisé en six domaines et une liste dont les intitulés lui ont donné son nom : Peuples et cultures, Anthroponymes, Chronologie, Toponymes (actuellement retirés), Œuvres, Lieux et Sujets. Adapté à l'indexation des collections des bibliothèques du réseau, il couvre désormais toutes les thématiques de l'archéologie, depuis la Préhistoire jusqu'à l'époque contemporaine, avec l'intégration de nouvelles ressources sur des périodes plus anciennes (Pré- et Protohistoire) ou plus récentes (du Moyen Âge à l'époque contemporaine) que l'Antiquité. Son assiette documentaire reprend un vocabulaire libre, hiérarchisé et interopérable, ce qui correspond à la philosophie d'ouverture des données adoptée par Bibracte. Une nouvelle version du thésaurus est en cours de consolidation pour en affiner l'organisation (https://pactols2.frantiq.fr/pactols2/).

[7] Créé en 2005 à la demande de la Fédération et ressources sur l'Antiquité (GDS *Frantiq*) pour la gestion des PACTOLS et développé sous la direction de Miled Rousset au sein de la Maison de l'Orient et de la Méditerranée Jean Pouilloux, *Opentheso* se positionne aujourd'hui comme un outil générique (https://thesaurus.mom.fr/opentheso/). Il est conforme aux normes ISO 25964-1 et ISO 25964-2.

[8] La norme internationale ISO 25964 est publiée en deux parties : ISO 25964-1 / Thésaurus pour la recherche d'informations (publiée en 2011 : https://www.iso.org/fr/standard/53657.html) et ISO 25964-2 / Interopérabilité avec d'autres vocabulaires (publiée en 2013 : https://www.iso.org/fr/standard/53658.html). L'ISO 25964-1 donne des recommandations pour le développement et la maintenance de thésaurus destinés à des applications de recherche d'information. Elle s'applique aux vocabulaires utilisés pour la recherche sur tous les types de ressources, quel que soit le support utilisé (texte, son, image fixe ou animée, objet physique ou multimédia), y compris les bases de connaissances et les portails, les bases de données bibliographiques, les collections de textes, de musées ou de multimédias, et les éléments qu'elles contiennent. Elle fournit également un modèle de données et un format recommandé pour l'importation et l'exportation de données de thésaurus. L'ISO 25964-2, qui s'applique aux thésaurus et autres types de vocabulaires utilisés pour la recherche d'informations, décrit, compare et confronte les éléments et les caractéristiques de ces vocabulaires à appliquer lorsque l'interopérabilité est nécessaire. Elle donne des recommandations pour l'établissement et la maintenance d'alignements ou *mappings* entre plusieurs thésaurus, ou entre des thésaurus et d'autres types de vocabulaires.

**L'application de la norme ISO 25964-1 pour l'archéologie**

L'archéologue a besoin d'exposer la singularité et la richesse de son espace sémantique, mais aussi de pouvoir accéder aux raisonnements de ses pairs par le biais d'un vocabulaire allant du plus général au plus précis selon les cas[9]. Le documentaliste, quant à lui, doit rendre compte par l'indexation de l'enrichissement constant des connaissances, ce qui l'oblige à compléter régulièrement le vocabulaire servant à l'indexation, avec le souci permanent de sélectionner et d'organiser les termes de signalement les plus pertinents en évitant d'empiler des vocabulaires/espaces sémantiques spécifiques, et en en produisant, au contraire, une synthèse. Ces deux approches sont essentielles et complémentaires, mais leur articulation a été peu envisagée jusqu'à présent.

Pour participer à l'enrichissement du vocabulaire d'indexation de PACTOLS, un prototype d'arborescence explicitant les besoins sémantiques spécifiques de l'archéologue a été formalisé et décrit dans un thésaurus nommé *Bibracte_Thesaurus* (http://ark.mom.fr/ark:/39676/srvtxcg5zrhv8). Cette expérimentation a permis de comprendre que les deux approches – celle de l'archéologue et celle du documentaliste – sont irréductibles l'une à l'autre, en raison d'une intentionnalité d'usage différente de la norme ISO 25964-1 : pour rendre compte du vocabulaire de l'archéologue, il a fallu recourir à une lecture adaptée de la norme ISO 25964-1 où le *concept* placé au sein d'une arborescence est la somme d'un *identifiant*, d'une *définition* et d'un *terme préféré* :

- L'identifiant (usuellement nommé URI, pour *Uniform Resource Identifier* (https://www.w3.org/Addressing)) est une courte chaîne de caractères permettant d'identifier une ressource de manière permanente sur un réseau physique ou abstrait, et dont la syntaxe respecte une norme d'Internet mise en place pour le *World Wide Web*. Dans *Opentheso*, l'URI est construit en partie sur l'IdArk (*Archival Resource Key*), un format facilitant l'identification de tous types de ressources (physiques, numériques ou même immatérielles) et dont le but est de fournir des identifiants uniques, pérennes et adaptés aux besoins des producteurs et diffuseurs de données sur le web.
- La définition est la concrétion d'indices convergents permettant à l'archéologue de préciser la notion qu'exprime le concept, à partir de plusieurs critères intrinsèques induits par son objet d'étude, son expérience, sa problématique et son contexte. Ces indices sont des données archéologiques (tableaux d'inventaire, analyses physico-chimiques, mesures, relevés planimétriques, etc.), des connaissances mobilisées par les chercheurs ou encore des représentations/modélisations (icono)graphiques (dessins, photographies, schémas, plans, etc.) de ce corpus. Ces données peuvent être synthétisées directement dans la définition et/ou être atteintes sous forme de ressources externes. La définition est essentielle et doit être sourcée (généralement par une référence bibliographique) pour être contextualisée.
- Le terme préféré (en langage Skos : *prefLabel*) est utilisé pour se référer de manière non ambiguë à une définition. C'est le descripteur d'une indexation et le libellé qui apparaît à l'affichage. C'est lui qui permet l'alignement vers des thésaurus externes. Le *synonyme* (en langage Skos : *altLabel*) est une variante du terme préféré.

Un exemple d'application est détaillé dans les pages qui suivent.

Dans cette approche, c'est la définition associée au terme préféré qui permet de différencier et de hiérarchiser les concepts pour rendre compte du raisonnement par deux types de relations :

- Les *relations hiérarchiques* se répartissent entre *concept générique* (niveau supérieur du *concept* consulté) et *concept spécifique* (niveau inférieur du concept consulté).
- Les *relations associatives* permettent de réunir des concepts qui ne sont pas liés hiérarchiquement dans une arborescence, mais dont la mise en commun est pertinente. Pour associer deux concepts, la norme veut que l'un soit un élément nécessaire à la définition de l'autre. Les relations associatives sont réciproques.

L'ensemble de ces relations forme un graphe répertoriant et contextualisant des ensembles de concepts, qu'ils soient publiés ou usuellement employés par les archéologues (référentiel archéologique), et traduisant un état du savoir, par nature situé dans le temps/millésimé, dans une branche du *Bibracte_Thesaurus*.

---

[9]   Une autre tentative de structuration selon la norme ISO 25964 d'un vocabulaire de la recherche (mobiliers métalliques : https://thesaurus.mom.fr/opentheso/api/theso/Artefacts) a été réalisée dans le cadre du projet *ArteBib* (2019-2020), qui prévoyait la mise à disposition sur la base de données en ligne collaborative Artefacts (https://artefacts.mom.fr/fr/home.php) de l'inventaire des objets archéologiques issus des fouilles anciennes et récentes réalisées sur l'oppidum de Bibracte.

**Formalisation du thésaurus : l'exemple de la céramique de Bibracte**

La publication de référence relative à l'étude de la céramique du site archéologique de Bibracte (Barrier, Luginbühl 2021) utilise un minimum de cinq concepts pour rendre compte dans la norme ISO 25964-1 des aspects mobilisés par les céramologues pour produire une description complète d'un artefact, par exemple un tesson d'assiette A15 en pâte grise fine et surface lissée fumigée (PGFINLF). Quatre d'entre eux classent l'artefact selon des critères spécifiques définis (forme, type, catégorie, chronologie) dont la concaténation forme la description chrono-morphologique de l'objet. Le cinquième concept source les définitions de ces critères spécifiques en renvoyant à un référentiel millésimé (référence bibliographique ou numérique) (Fig. 1 ; Fig. 2).

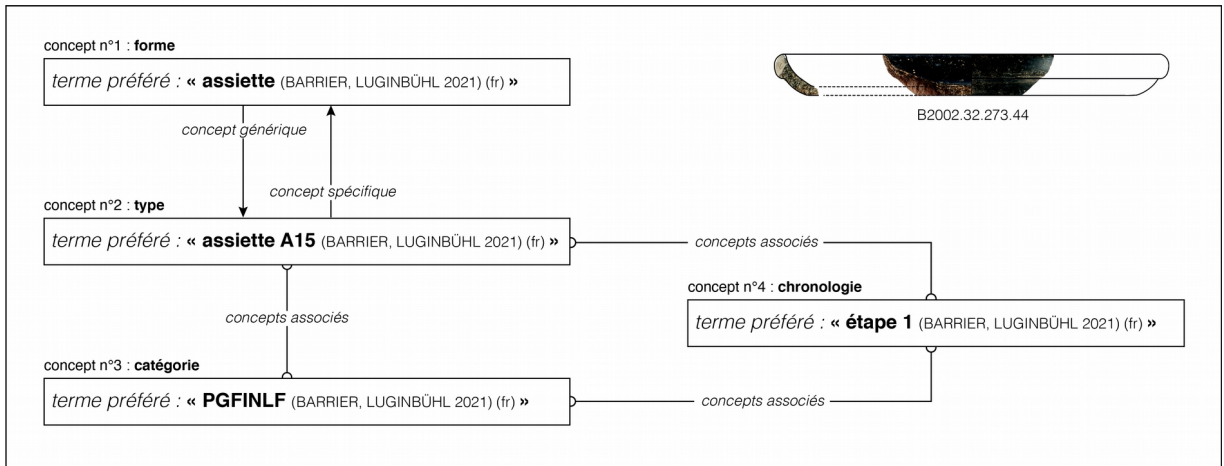

*Fig. 1: Description morpho-chronologique du tesson B2002.32.273.44.*

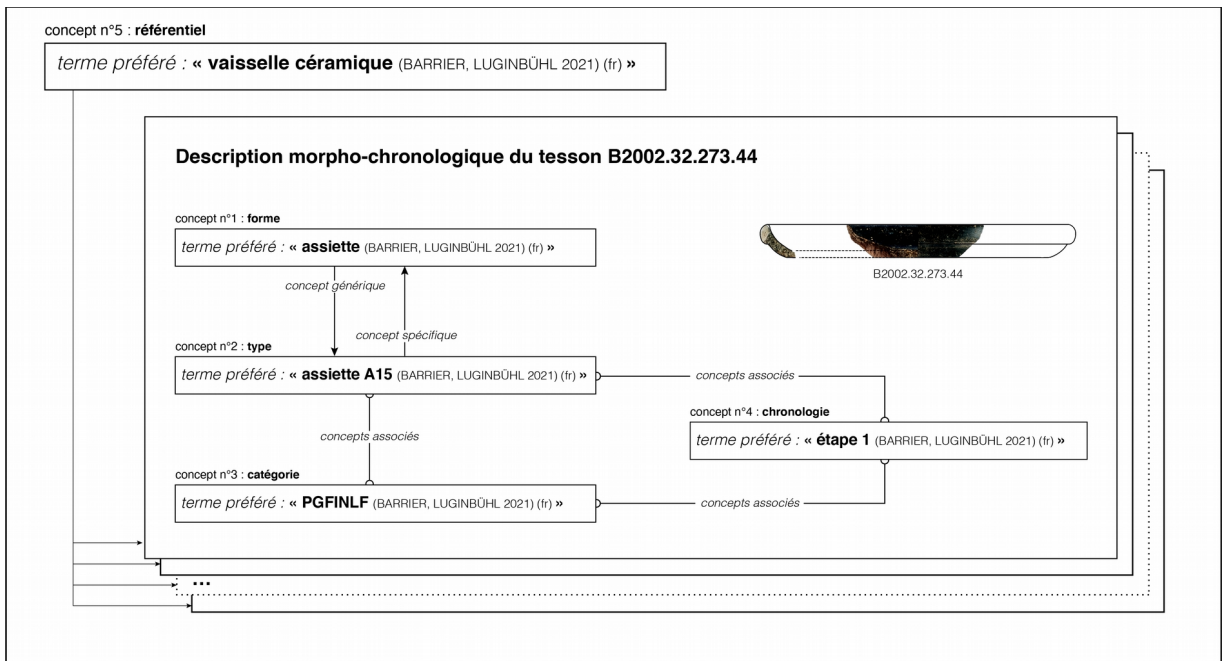

*Fig. 2: Contextualisation du vocabulaire d'un référentiel millésimé.*

Les cinq concepts mobilisés sont construits et contextualisés de la façon suivante :

**Concept n°1 : forme**
URI / Identifiant IdArk : https://ark.mom.fr/ark:/39676/bib25gwqwnprh
Définition : « Forme basse, ouverte, de hauteur inférieure au quart du diamètre. Diamètre compris entre 15 et 25 cm. Pied annulaire. Catégories : importations, céramiques de tradition méditerranéenne et céramiques fines régionales. Origine culturelle : méditerranéenne. Fonctions présumées : servir et présenter les aliments. (Source : Barrier, Luginbühl 2021) »
Ressource externe :
https://api.nakala.fr/data/10.34847/nkl.89b20d19/db80f98eb18efa07e04bc7287de8926fa5e8cca9
Terme préféré : « assiette (BARRIER, LUGINBÜHL 2021) (fr) »
0 synonyme
1 relation hiérarchique :
- 1 concept générique : « formes (BARRIER, LUGINBÜHL 2021) (fr) »
- 0 concept spécifique : aucun (au bout de l'arborescence)
Relations associatives : « types assiettes (BARRIER, LUGINBÜHL 2021) (fr) »
Chemin d'accès (arborescence) : Bibracte_Thesaurus > 3 - mobilier > artefacts > céramique (mobilier) > récipients en céramique > céramique période oppidum > vaisselle période oppidum > vaisselle (BARRIER, LUGINBÜHL 2021) > formes (BARRIER, LUGINBÜHL 2021) > assiette (BARRIER, LUGINBÜHL 2021)

**Concept n°2 : type**
URI / Identifiant IdArk : https://ark.mom.fr/ark:/39676/bibxtjgnrpkr5
Définition : « Plat à paroi concave, lèvre arrondie épaissie en bandeau. Imitation de R-Pomp 1 (plat à engobe interne italien). (Source : Barrier, Luginbühl 2021). »
Ressource externe :
https://api.nakala.fr/data/10.34847/nkl.89b20d19/b6f784626ed44f0443a5fa62e78b7759a0b1a4f6
Terme préféré : « Assiette A15 (BARRIER, LUGINBÜHL 2021) (fr) »
3 synonymes : « A15 », « plat A15 », « plat à paroi concave et lèvre arrondie épaissie en bandeau »
1 relation hiérarchique :
- 1 concept générique : « types assiettes (BARRIER, LUGINBÜHL 2021) (fr) »
- 0 concept spécifique : aucun (au bout de l'arborescence)
3 relations associatives : « EIRA (BARRIER, LUGINBÜHL 2021) (fr) », « PGFINLF (BARRIER, LUGINBÜHL 2021) (fr) », « PGFINTN (BARRIER, LUGINBÜHL 2021) (fr) »
Chemin d'accès (arborescence) : Bibracte_Thesaurus > 3 - mobilier > artefacts > céramique (mobilier) > récipients en céramique > céramique période oppidum > vaisselle période oppidum > vaisselle (BARRIER, LUGINBÜHL 2021) > types (BARRIER, LUGINBÜHL 2021) > types assiettes (BARRIER, LUGINBÜHL 2021) > assiette A15 (BARRIER, LUGINBÜHL 2021)

**Concept n°3 : catégorie**
URI / Identifiant IdArk : https://ark.mom.fr/ark:/39676/bibrbqbp0019d
Définition : « Céramique à pâte grise fine et surface lissée fumigée. Surface : Parois lissées, assez ou peu luisantes (sauf intérieur des formes fermées), gris foncé ou noires. Types de décors variés (imprimés, imprimés à la molette, polis). Pâte : Siliceuse, fine, dure, gris moyen. Montage : Tournage et tournassage. Répertoire : Vaisselle de table et de stockage d'appoint. Assiettes, plats, écuelles, coupes, bols, gobelets, pots, bouteilles, tonnelets, rares cruches. Origine : Régionale. Chronologie : Production probable dès avant 120 av. n. è. jusqu'au début du Ier s. de n. è. (Source : Barrier, Luginbühl 2021) »
Ressource externe :
https://api.nakala.fr/data/10.34847/nkl.89b20d19/07095a124e9f88122f9b4b064a7c728fd580cf45
https://api.nakala.fr/data/10.34847/nkl.89b20d19/1aeee1d1e44acbbb6de6d17aebe1cb7131eb0d7e
Terme préféré : « PGFINLF (BARRIER, LUGINBÜHL 2021) (fr) »
1 synonyme : « céramique à pâte grise fine et surface lissée fumigée »
1 relation hiérarchique :
- 1 concept générique : « céramique tournée lissée à pâte sombre/grise (BARRIER, LUGINBÜHL 2021) (fr) »
- 0 concept spécifique : aucun (au bout de l'arborescence)
197 relations associatives : « assiette A1 (BARRIER, LUGINBÜHL 2021) (fr) », « assiette A10 (BARRIER, LUGINBÜHL 2021) (fr) », « assiette A10a (BARRIER, LUGINBÜHL 2021) (fr) », « assiette A10b (BARRIER, LUGINBÜHL 2021) (fr) », « assiette A11 (BARRIER, LUGINBÜHL 2021) (fr) », « assiette A11a (BARRIER, LUGINBÜHL 2021) (fr) », « assiette A11b (BARRIER, LUGINBÜHL 2021) (fr) », « assiette A15 (BARRIER,

LUGINBÜHL 2021) (fr) » […] « types tonnelets (BARRIER, LUGINBÜHL 2021) (fr) », « types vases bobine (BARRIER, LUGINBÜHL 2021) (fr) », « [céramique à pâte grise] (fr) »

Chemin d'accès (arborescence) :  Bibracte_Thesaurus > 3 - mobilier > artefacts > céramique (mobilier) > récipients en céramique > céramique période oppidum > vaisselle période oppidum > vaisselle (BARRIER, LUGINBÜHL 2021) > catégories (BARRIER, LUGINBÜHL 2021) > céramique tournée (BARRIER, LUGINBÜHL 2021) > céramique tournée lissée (BARRIER, LUGINBÜHL 2021) > céramique tournée lissée à pâte sombre/grise (BARRIER, LUGINBÜHL 2021) > PGFINLF (BARRIER, LUGINBÜHL 2021)

**Concept n°4 : chronologie**
URI / Identifiant IdArk : https://ark.mom.fr/ark:/39676/bib2q5s0bw54c
Définition : « Marqueurs de la céramique :
- Importations méditerranéennes principalement représentées par des campaniennes A et B, ainsi que des gobelets italiens à parois fines (Mayet I et II). Rares bols hellénistiques à reliefs.
- Productions d'influence méditerranéenne constituées de cruches à col large (Gaule méridionale ou rhodanienne) et d'imitations de pichets de type ampuritain (Auvergne septentrionale ?).
- Faciès des fines régionales caractérisé par des bouteilles peintes à fond blanc et décor de cervidés, des tonnelets peints à fond lie-de-vin, parfois orné de pastilles en réserve, ainsi que des productions « grises fines » diversifiées (groupes à surface brune, à cœur rouge, à surface lustrée et à surface lissée fumigée).
- Productions mi-fines encore rares (proportions aux alentours de 10 %), alors que les grossières constituent entre environ 40 % et 50 % des ensembles. (Source : Barrier, Luginbühl 2021) »
Ressource externe : /
Terme préféré : « Étape 1 céramique : 120/110 à 90/80 av. n.è. (BARRIER, LUGINBÜHL 2021) (fr) »
0 synonyme
1 relation hiérarchique :
- 1 concept générique : « périodisation BARRIER, LUGINBÜHL 2021 (fr) »
- 0 concept spécifique : aucun (au bout de l'arborescence)
27 relations associatives : « CAMPA (BARRIER, LUGINBÜHL 2021) (fr) », « CAMPB (BARRIER, LUGINBÜHL 2021) (fr) », « MICACB (BARRIER, LUGINBÜHL 2021) (fr) », « MICACBCN (BARRIER, LUGINBÜHL 2021) (fr) », « MICACFIN (BARRIER, LUGINBÜHL 2021) (fr) », « MICACG (BARRIER, LUGINBÜHL 2021) (fr) », « MICACGCN (BARRIER, LUGINBÜHL 2021) (fr) », « MODGROS (BARRIER, LUGINBÜHL 2021) (fr) », « PARFINA (BARRIER, LUGINBÜHL 2021) (fr) », « PARFINC (BARRIER, LUGINBÜHL 2021) (fr) », « PCCRU (BARRIER, LUGINBÜHL 2021) (fr) », « PCCRUENG (BARRIER, LUGINBÜHL 2021) (fr) », « PCGROS (BARRIER, LUGINBÜHL 2021) (fr) », « PCGROSCN (BARRIER, LUGINBÜHL 2021) (fr) », « PCLUSTR (BARRIER, LUGINBÜHL 2021) (fr) », « PCMIFIN (BARRIER, LUGINBÜHL 2021) (fr) », « PEINTA (BARRIER, LUGINBÜHL 2021) (fr) », « PEINTB (BARRIER, LUGINBÜHL 2021) (fr) », « PEINTC (BARRIER, LUGINBÜHL 2021) (fr) », « PGCAT (BARRIER, LUGINBÜHL 2021) (fr) », « PGFINLF (BARRIER, LUGINBÜHL 2021) (fr) », « PGLUSTR (BARRIER, LUGINBÜHL 2021) (fr) », « PGMIFIN (BARRIER, LUGINBÜHL 2021) (fr) », « PSFINA (BARRIER, LUGINBÜHL 2021) (fr) », « PSFINB (BARRIER, LUGINBÜHL 2021) (fr) », « PSGROS (BARRIER, LUGINBÜHL 2021) (fr) », « VRHELLEN (BARRIER, LUGINBÜHL 2021) (fr) »
Chemin d'accès :  Bibracte_Thesaurus > 4 - chronologie > périodisation BARRIER, LUGINBÜHL 2021 > Étape 1 céramique : 120/110 à 90/80 av. n.è. (BARRIER, LUGINBÜHL 2021)

**Concept n°5 : référentiel**
URI / Identifiant IdArk : https://ark.mom.fr/ark:/39676/bibd9q291x45d
Définition : « Ce référentiel est dérivé de la publication suivante : BARRIER (S.), LUGINBÜHL (T.), BARRAL (P.) coll. — La vaisselle céramique de Bibracte. De l'identification à l'analyse. Glux-en-Glenne : Bibracte, 2021. (Bibracte, 31 ; ISSN : 1281-430X ; ISBN : 978-2-490601-07-3), 318 pages, 177 illustrations.
Premier élément date et référence bibliographique : Barrier, Luginbühl 2021 : Barrier (S.), Luginbühl (T.) — La vaisselle céramique de Bibracte. De l'identification à l'analyse. Glux-en-Glenne : Bibracte, 2021. 318 p., 177 ill. (Bibracte ; 31).
Mots-clefs de l'ouvrage (termes sélectionnés sur PACTOLS 2, thésaurus accessible en 2022) : Celtes ; Eduens ; Bibracte ; céramique (matériau) ; vaisselle ; La Tène ; IIe siècle av. J.-C. ; Ier siècle av. J.-C. ; Ier siècle ; récipient ; vie quotidienne ; alimentation ; artisanat »
Ressource externe : /
Terme préféré : « vaisselle céramique (BARRIER, LUGINBÜHL 2021) (fr) »
0 synonyme
2 relations hiérarchiques :

- 1 concept générique : « 5 - référentiels (fr) »
- 1 concept spécifique : « bibliographie (BARRIER, LUGINBÜHL 2021) (fr) »
3 relations associatives : « catégories (BARRIER, LUGINBÜHL 2021) (fr) », « périodisation BARRIER, LUGINBÜHL 2021) (fr) », « types (BARRIER, LUGINBÜHL 2021) (fr) », « vaisselle (BARRIER, LUGINBÜHL 2021) (fr) »
Chemin d'accès (arborescence) : Bibracte_Thesaurus > 5 - référentiels > vaisselle céramique (BARRIER, LUGINBÜHL 2021)

Le référentiel « vaisselle céramique (BARRIER, LUGINBÜHL 2021) (fr) », transposition numérique sous forme de thésaurus d'une partie du contenu d'une ressource bibliographique millésimée, permet l'accès libre et permanent à la définition et à la datation des 53 catégories, des 25 formes et des 423 types de céramiques répertoriés à ce jour sur le site pour la période d'occupation de l'*oppidum* de Bibracte, donnant un plafond théorique de 22.419 combinaisons (53 catégories * 423 types répartis en 25 formes). Il témoigne d'une réalité archéologique matérielle riche (plus de 300.000 éléments enregistrés dans la base de données du site) et génère des concepts de granularité fine, équilibrés selon une logique construite par l'usage. Il reflète la problématique initiale des auteurs : fournir les clés de l'identification d'un tesson grâce à l'observation et à l'expérience du céramologue, du plus spécifique (le tesson) au plus générique (un groupe de tessons partageant les mêmes propriétés), dans le cadre de l'élaboration d'une classification morphotypologique. Par la suite, dans un contexte informationnel, les concepts seront présentés du plus générique au plus spécifique : « céramique tournée (BARRIER, LUGINBÜHL 2021) (fr) » > « céramique tournée lissée (BARRIER, LUGINBÜHL 2021) (fr) > « céramique lissée tournée à pâte sombre/grise (BARRIER, LUGINBÜHL 2021) (fr) » > « PGFINLF (BARRIER, LUGINBÜHL 2021) (fr) » ; les définitions, construites sur une réalité matérielle tangible, garantissent la compréhension du libellé (*prefLabel* et *altLabel*) de chacun des concepts.

**Discussion**

La norme ISO 25964-1 fournit un cadre structurel utile à l'archéologue pour créer son vocabulaire, organiser une approche pédagogique de sa nomenclature et l'outiller pour la comparaison de sa terminologie avec les champs lexicaux développés par d'autres chercheurs. Les règles et les contraintes formelles de la norme, comme l'impossibilité de produire deux termes strictement identiques ou la hiérarchisation depuis une nomenclature générique vers un vocabulaire de fine granularité, deviennent des outils efficaces pour traquer et décrire les critères implicites du raisonnement et progresser rigoureusement vers une approche systématique et explicite. Par ailleurs, la vue arborescente du thésaurus sur *Opentheso*, des termes les plus génériques vers les concepts les plus spécifiques, favorise l'exploration du sujet, notamment lorsque l'archéologue cherche à s'imprégner des univers sémantiques et des paradigmes de ses collègues.

L'exemple de la céramique de Bibracte montre qu'il est possible de formaliser un vocabulaire « orienté recherche » dans la norme ISO 25964-1 en respectant les règles. Cependant, la spécificité de l'intentionnalité de ce vocabulaire (*i. e.* dans ce cas précis l'identification d'un tesson) a conduit à s'écarter de l'application usuelle (documentaire) de cette norme, en privilégiant la définition plutôt que les libellés (*prefLabel* et *altLabel*) d'un concept. Cette approche, dans laquelle le terme isolé (décontextualisé) n'a pas sa place, a pour conséquence logique l'évolutivité du thésaurus par l'enrichissement itératif de ses définitions, induisant potentiellement le déplacement du concept dans l'arborescence. L'outil thésaurus rend ici compte des connaissances produites par la recherche et nécessaires à l'activité scientifique. *A contrario*, les thésaurus documentaires singularisent un terme descripteur (*prefLabel*) et ses variantes (*altLabel*) pour obtenir un concept non-ambigu figé dans une arborescence stable répondant aux contraintes du multilinguisme et de l'accessibilité par tout usager. Mais sans sa définition, le concept est réduit à une simple « information », entendue ici à la fois comme facteur d'organisation et comme messager (Shannon, Weaver 1949, pp. 33-35) ; en contrepartie, la structuration tend alors vers la pérennité. Ces deux approches qui semblent inconciliables peuvent-elles se rejoindre dans le cadre d'un alignement entre thésaurus ?

Une tentative d'alignement entre les deux modèles organologiques a été opérée, en rapprochant via l'exemple de la céramique, le vocabulaire en usage à Bibracte avec le thésaurus d'indexation bibliographique PACTOLS 2. Cette tentative a rencontré plusieurs difficultés.

Ainsi, le concept du *Bibracte_Thesaurus* dont le *prefLabel* est « CAMPA (BARRIER, LUGINBÜHL 2021) (fr) » (avec pour *altLabel* « céramique à vernis noir campanienne A ») et dont la définition est « *Céramique à vernis* *noir campanienne A. Surface : Vernis noir, luisant (parfois légèrement métallescent), adhérant bien. Pâte : Calcaire, fine, dure, rose saumon. Montage : Tournage et tournassage. Répertoire : Vaisselle de table. Assiettes, plats, coupes et bols principalement. Origine : Campanie. Chronologie : Catégorie surtout caractéristique du IIe s. av. n. è., dont la production chute dès le début du siècle suivant. (Source : Barrier, Luginbühl 2021)* » trouve dans le thésaurus d'indexation bibliographique PACTOLS 2 un alignement avec le concept dont le *prefLabel* est « céramique campanienne A (fr) » (sans *altLabel* ni définition). Seules la définition du concept et la présence bienvenue d'un *altLabel* coïncidant avec le *prefLabel* de PACTOLS 2 autorisent à proposer un alignement entre ces deux libellés. L'alignement que nous avons effectué en *exactMatch* n'est pas satisfaisant, car il fait implicitement référence à plusieurs critères intrinsèques induits par l'objet d'étude et son contexte, l'expérience et la problématique de l'archéologue.

Par ailleurs, l'exemple d'un tesson d'assiette A15 en pâte grise fine et surface lissée fumigée (PGFINLF), pour lequel la description complète mobilise cinq concepts (forme, type, catégorie, chronologie et référentiel) dont la formulation linguistique est encore nourrie d'implicite, révèle le besoin d'un espace sémantique intermédiaire à l'interface entre un vocabulaire « orienté recherche » et un vocabulaire d'indexation bibliographique, car les deux modèles sont loin d'être systématiquement raccordables. En effet, le concept « assiette (BARRIER, LUGINBÜHL 2021) (fr) » peut être aligné avec le terme préféré « assiette (fr) » sur PACTOLS 2, mais la définition de ce concept y est plus générique (ou dissemblable) : « *Récipient à parois fortement évasées dont le diamètre à l'ouverture (inférieur ou égal à 23/24 cm environ) est égal ou supérieur à cinq fois la hauteur (Balfet* et al. *1989, p. 10). Récipient de forme générale très évasée avec un large marli et une base légèrement marquée, dont le diamètre d'ouverture est supérieur à 5 fois la hauteur (ICÉRAMM 2021)* ». Pourquoi ? Parce qu'elle s'appuie elle-même sur deux autres typologies d'auteurs, très proches sans être semblables à celle en usage à Bibracte (différences de diamètre ou le rapport diamètre/hauteur) ; on remarquera d'ailleurs que les deux typologies mobilisées ne disent pas exactement la même chose.

Dans ces deux cas apparemment opposés, c'est l'écart ou la tentative de résorber l'écart entre deux pratiques irréductibles l'une à l'autre qui empêche un alignement satisfaisant (*exactMatch*[10]) et suggère au contraire le recours à des relations hiérarchiques, où le concept du *Bibracte_Thesaurus* est un *broadMatch* du *prefLabel* de PACTOLS 2 (ce dernier devant idéalement émettre en retour un lien inverse en *narrowMatch*[11]). Le terme préférentiel « assiette A15 (BARRIER, LUGINBÜHL 2021) (fr) », qui ne permet aucun alignement pertinent (le raccord le moins frustrant pourrait se faire sur le *prefLabel* « assiette (fr) »), révèle quant à lui l'ampleur de l'espace sémantique intermédiaire manquant, car c'est dans la granularité fine des détails que réside l'essentiel de l'information typochronologique pour l'archéologue.

La catégorie « PGFINLF (BARRIER, LUGINBÜHL 2021) (fr) » a, quant à elle, nécessité la création artificielle d'un terme de regroupement dans le *Bibracte_Thesaurus*, en l'occurrence « [céramique à pâte grise] » – qui ne renvoie à aucune notion définie explicitement par les producteurs du référentiel « vaisselle céramique (BARRIER, LUGINBÜHL 2021) (fr) » – pour autoriser un lien vers le concept générique « céramique à pâte grise (fr) » sur PACTOLS 2. Dans le même registre, la portée chronologique contenue dans le concept « Étape 1 céramique : 120/110 à 90/80 av. n.è. (BARRIER, LUGINBÜHL 2021) (fr) » ne trouve aucun écho sur le thésaurus PACTOLS 2, sauf à envisager un alignement sur « IIe siècle av. J.-C. (fr) » ou « Ier siècle av. J.-C. (fr) » qui ne renvoient pas à la même réalité.

Les obstacles en matière d'alignement entre les deux systèmes de pensée et de pratique soulignent le besoin d'un espace sémantique faisant office de « pivot », susceptible de générer une forme de consensus. Ce qui semble n'être qu'un impératif contraignant dicté par les besoins de l'alignement, à savoir la création de termes de regroupement, pourrait finalement devenir le début d'une méthode bénéfique d'enrichissement des thésaurus archéologiques de haut niveau fondée sur une réalité matérielle plus proche du terrain. La constitution d'un tel pivot soulève bien sûr des questions méthodologiques, notamment par son impact sur le choix, la définition et la structuration des concepts. Ce niveau de normalisation sémantique doit être construit dans un dispositif de coopération régulée qui associe un ou des locuteurs de la langue du domaine, spécialistes-experts (en l'espèce : des archéologues), et un ingénieur de la connaissance maîtrisant l'usage de la norme, dont le rôle n'est pas celui d'arbitrer des choix mais d'être un facilitateur méthodologique de l'élaboration d'une sémantique réellement collégiale (Bachimont 2000, p. 6) (Fig. 3).

---

[10]   Équivalence exacte entre deux concepts.

[11]   Dans l'exemple où A est un concept générique et B est un concept spécifique, le concept A permet de créer une équivalence de type *narrowMatch* vers le concept B, tandis que le concept B permet de créer une équivalence de type *broadMatch* vers le concept A.

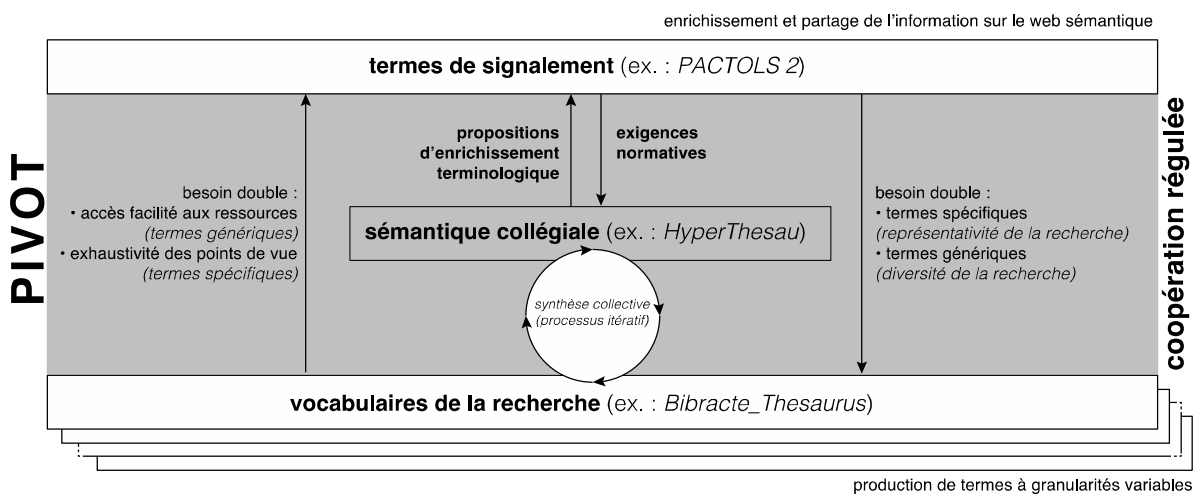

*Fig. 3: L'élaboration d'une sémantique collégiale entre besoins et contraintes heuristiques et documentaires.*

Pour se rapprocher des alignements exacts (*exactMatch*) qui sont aujourd'hui la condition *sine qua non* d'une interopérabilité machine à machine entre ces deux familles de thésaurus, la voie choisie par *HyperThésau* consiste à contribuer à la construction de ce thésaurus-pivot (https://thesaurus.mom.fr/), conçu comme un outil de médiation entre des vocabulaires « locaux » ou idiolectes scientifiques et les vocabulaires plus généraux des référentiels pérennes, extérieurs à la communauté disciplinaire mais ouverts sur le web sémantique, dotés des moyens et d'une autorité compatibles avec leur vocation universelle (Perrin *et al.* 2020, p. 2).

Cette voie n'est pas sans soulever un redoutable défi : l'espace d'incertitude inhérent à la prise en compte de ces « points de vue » scientifiques proches et néanmoins distincts. Un précédent inspirant existe néanmoins : le référentiel *PeriodO* (https://perio.do/en/) organise, non pas des concepts de « périodes » *in abstracto*, mais des définitions sourcées de périodes associées à un intervalle de dates et à une zone géographique précis par un travail de recherche identifié, ce qui rend cette incertitude calculable et permet l'organisation automatisée des « points de vue » chronologiques. La prise en compte d'incertitudes linguistiques supposera sans doute de les rendre pareillement calculables, donc de faire appel, peu ou prou, à des outils d'intelligence artificielle (IA) appliqués tant à la structuration des concepts (graphes) qu'au contenu sémantique de leurs définitions et à l'analyse des objets graphiques associés.

L'apport bénéfique de l'IA n'est envisageable qu'à condition de constituer les réservoirs de données qui fournissent son carburant. Pour cela, trouver, accéder, interopérer et réutiliser les données archéologiques dans ce réservoir est le premier défi, que l'on ne peut relever qu'à la condition de rendre cette opération accessible en mettant au point une méthode que tous les archéologues pourront s'approprier. Nous présenterons cette méthode fondée sur le vocabulaire dans le deuxième article de cette série.

**Bibliographie**